\documentclass[journal=jacsat,manuscript=article]{achemso}

\usepackage{chemformula} 
\usepackage[T1]{fontenc} 
\usepackage{multicol}
\usepackage{braket}
\usepackage{amsmath}
\usepackage{amssymb}
\usepackage{cancel}
\usepackage{hyperref}
\usepackage{xcolor}
\hypersetup{
    colorlinks,
    linkcolor={red!50!black},
    citecolor={blue!50!black},
    urlcolor={blue!80!black}
}

\author{Luke Cheeseman}
\author{Kayn A. Forbes}
\affiliation[BigPharma]
{School of Chemistry, University of East Anglia, Norwich, Norfolk, NR4 7TJ, United Kingdom}
\email{k.forbes@uea.ac.uk}
\title[An \textsf{achemso} demo]
  {Nonlinear vortex dichroism in chiral molecules}

\abbreviations{IR,NMR,UV}

\begin{document}


\begin{abstract}
The recent discovery that linearly polarized light with a helical wavefront can exhibit vortex dichroism (also referred to as helical dichroism) has opened up new horizons in chiroptical spectroscopy with structured chiral light. Recent experiments have now pushed optical activity with vortex beams into the regime of nonlinear optics. Here we present the theory of two-photon absorption (TPA) of focused optical vortices by chiral molecules: nonlinear vortex dichroism (NVD). We discover that highly distinct features arise in the case of TPA with focused vortex beams, including the ability to probe chiral molecular structure not accessible to current methods and that the differential rate of TPA is significantly influenced by the orientation of the state of linear polarization. This work provides strong evidence that combining nonlinear optical activity with structured light provides new and improved routes to studying molecular chirality. 

\end{abstract}

\section{Introduction}
An object is chiral if it cannot be superimposed onto its mirror image. In the microscopic world, a molecule is chiral if it lacks any inversion, mirror reflection or
other rotation-reflection symmetry elements in its relevant point group, coming in left- and right-handed forms known as enantiomers. Light may also be chiral (see Figure 1). The simplest and most known example of this is the helical rotation of the electric field vector of a circularly polarized plane wave: it can twist to the left $\sigma = 1$ or to the right $\sigma = -1$ as it propagates through space ($\sigma$ is often referred to as the helicity). It is well established that chiral molecules and nanostructures show a differential response towards the handedness of circularly polarized light in light-matter interactions: this discrimination is known as natural optical activity \cite{barron2009molecular, craig1998molecular}. Interacting chiral molecules with chiral light underpins chiroptical spectroscopy \cite{barron2009molecular, berova2011comprehensive, polavarapu2016chiroptical} and allows for the determination of the absolute configuration of chiral molecules using purely optical methods. Given that almost all biological molecules are chiral and over 50\% of pharmaceutical drugs are too, the importance of being able to determine absolute configuration of chiral molecules cannot be overstated. The use of circularly polarized light to discriminate enantiomers is routinely carried out in experiments based on absorption (dichroism), scattering (e.g. Raman and Rayleigh), and optical rotation. Such chiral spectroscopy, or chiroptical spectroscopy, has grown into a very well established research field with widespread applications in chemical systems, \cite{krupova2020recent,leung2012rapid,barron2009molecular} biomolecules \cite{Kumar2018detection,Woody2012comprehensive,Santiago2022bioplasmonics,Micsonai2015secondary, kakkar2020superchiral, kelly2018chiral, Kakkanattu2021sensing, chaubey2024ultrasensitive}, and artificial plasmonic nanostructures and metamaterials \cite{Goerlitzer2021plasmonics,Kakkanattu2021sensing,Chen2022nanoscopic,Zhang2021sensing,Kong2020plasmonic, kartau2023chiral, collins2017chirality}. 

\begin{figure*}
    \includegraphics[]{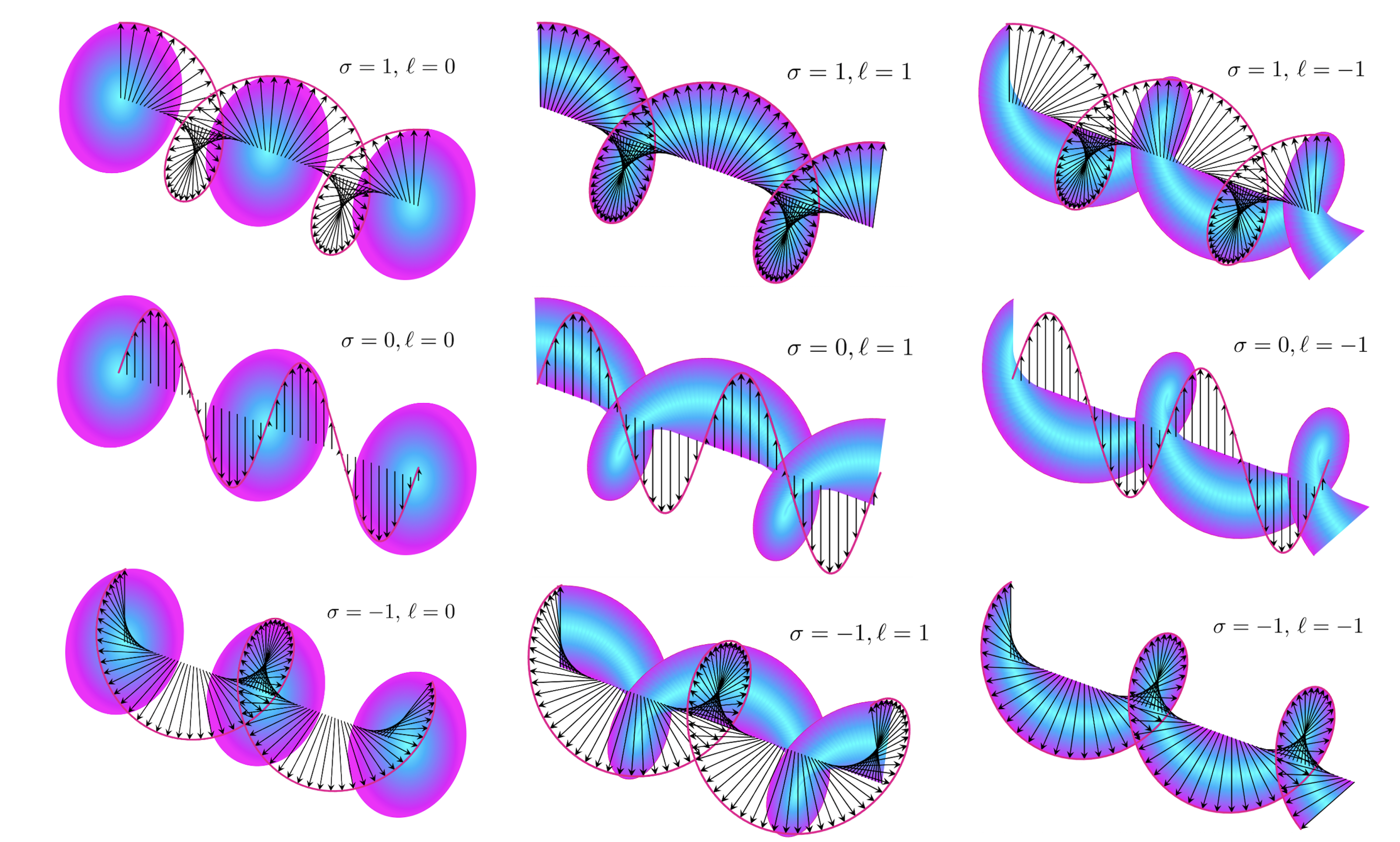}
    \caption{Chiral structures of polarization (arrows) and wavefront (coloured surface of constant phase) in paraxial electromagnetic beams: left-hand column Gaussian beam ($\ell = 0)$; middle and right-hand column Laguerre-Gaussian optical vortex $\ell \neq 0$. The sign of two pseudoscalars $\sigma$ and $\ell$ determine the handedness of the polarization state and wavefront, respectively. In the optics convention $\sigma, \ell > 0$ are left-handed and $\sigma, \ell <0$ right-handed. For linearly-polarized light $\sigma = 0$ and the polarization structure is not chiral. As can be seen from the left-hand column, Gaussian beams have a chiral structure due to a non-zero value of $\sigma$, i.e. a degree of ellipticity, being maximum for circular polarization $\sigma \pm 1$, but their wavefront is achiral. The middle and right-hand columns highlight the much richer possibilities with optical vortex beams with their chiral wavefront, including the fact such beams are still chiral even when linearly-polarized and also the distinct interplay between wavefront and polarization handedness, e.g. $\sigma = 1, \ell =1$ has a completely different structure to $\sigma = -1, \ell = 1$}.
    \label{fig:1}
\end{figure*}

Beyond polarization, chirality can be also manifest in other degrees of freedom of light such as the spatial distribution of phase in structured light beams. A Laguerre-Gaussian (LG) optical vortex mode is a common example of a structured light beam with a helical wavefront (surface of constant phase) due to its azimuthal phase $\text{e}^{i\ell\phi}$, where $\phi$ denotes the azimuthal angle around the beam profile and $\ell$ is the topological charge, a pseudoscalar which can take on both positive and negative integer values (see Figure \ref{fig:1}). For $\ell>0$ the helical wavefront twists to the left, $\ell<0$ wavefronts twist to the right ($\ell=0$ corresponds to a Gaussian mode). Since pioneering work in the early 90s showing these beams carry orbital angular momentum (OAM) of $\ell \hbar$ per photon, optical vortices have been extensively studied predominantly due to their diversity of application across distinct areas such as optical manipulation, microscopy, optical communications, and quantum information to name a few \cite{andrews2011structured, shen2019optical, forbes2021structured}.

An emerging area of research that brings together chiroptical spectroscopy and structured light uses the chirality associated with the wavefront handedness (sign of $\ell$) of optical vortices to produce optical activity. The motivation for this is twofold: firstly, compared to circularly polarized light $\sigma = \pm1$, the topological charge can take on arbitrarily large values $\ell \in \mathbb{Z}$, in principle leading to larger chiral light-matter interactions; and secondly, chiral effects can manifest without the need for circular polarization (in fact even unpolarized optical vortices can produce optical activity \cite{forbes2022optical}). Crucial to being able to engage the chirality of optical vortices in light-matter interactions is a beam which is tightly focused, bringing the chirality of the wavefront (a global property of the beam which scales with the beam waist) to similar dimensions as the chiral material being probed. Spatial confinement of the vortex beam through focusing leads to a decreasing  beam waist to wavelength ratio, a breakdown of the paraxial approximation, and the generation of significant longitudinal electromagnetic fields (electromagnetic field components parallel to the direction of propagation) which give rise to the associated optical activity \cite{forbes2021measures, green2023optical}. Such beam focusing is a common feature of laser microscale analysis.

Numerous studies have recently emerged showcasing the ability for vortex chirality to probe material chirality (see reviews \cite{forbes2021orbital, porfirev2023light}). Novel forms of optical activity depending on the topological charge include vortex dichroism (VD) during one-photon absorption \cite{wozniak2019interaction, forbes2021optical} -  note, this is sometimes referred to as helical dichroism (HD) by some authors - and vortex-dependent differential scattering \cite{forbes2019raman, ni2021gigantic, mullner2022discrimination}. Such studies have highlighted the capability of structured light to produce more sensitive and incisive techniques in chiral spectroscopy, sparking a surge in research activity aimed at expanding the range of new chiroptical techniques exploiting vortex chirality \cite{forbes2021orbital, porfirev2023light}. Cutting edge experiments have broken into the nonlinear regime, involving vortex-dependent nonlinear chiroptical methods studying organometallic complexes, crystalline solids, and enantiomeric solutions \cite{rouxel2022hard, begin2023nonlinear,jain2023helical,jain2024intrinsic}. Inspired by these experiments, here we detail the theory of nonlinear VD (i.e. two-photon absorption (TPA)) of tightly-focused LG beams by an isotropic assembly of chiral molecules. We discover highly distinct and novel features of nonlinear optical activity using the chirality of structured light, including vortex-dependent absorption for input linearly-polarized beams, access to structural information not available to standard techniques, and the advantages provided by tailorable excitation sources.

\section{Two-photon absorption rate for chiral molecules irradiated by a focused vortex beam}

In the theory of molecular quantum electrodynamics (MQED) \cite{craig1998molecular,salam2009molecular,andrews2020quantum, woolley2022foundations} the coupling between light and molecules may be represented by the Power-Zienau-Woolley (PZW) interaction Hamiltonian $H_\text{int}$, which for a molecule $\xi$ positioned at $\mathbf{R}_{\xi}$ is given by: 

\begin{align}
    H_{\mathrm{int}}(\xi) =   - \boldsymbol{\mu}{(\xi)} \cdot \mathbf{E}(\mathbf{R}_\xi) - \mathbf{m}{(\xi)} \cdot \mathbf{B}(\mathbf{R}_\xi) +... \mathrm{h.o.t.},
\label{eq:1}
\end{align}

where $\boldsymbol{{\mu}}(\xi)(\mathbf{m}(\xi))$ is the electric (magnetic) transition dipole moment operator and $\mathbf{E}(\mathbf{R}_\xi)(\mathbf{B}(\mathbf{R}_\xi))$ is the electric (magnetic) field operator evaluated at $\mathbf{R}_{\xi}$. It is well-established that natural optical activity manifests through the interferences between electric-dipole coupling (E1) with magnetic-dipole (M1) and electric quadrupole (E2), E1M1 and E1E2, respectively \cite{barron2009molecular, craig1998molecular}. Furthermore, in an isotropic (randomly oriented) ensemble the E1E2 contributions vanish in both one-photon and two-photon dichroism \cite{power1975two}, and so in this paper we concentrate solely on the E1M1 mechanism. Nevertheless, E1E2 terms do contribute in oriented or anisotropic chiral materials and these contributions to nonlinear VD are calculated and discussed in the Supplementary Information \cite{supplement}. The electric field and magnetic field mode expansions for an LG beam propagating along $z$, are given in cylindrical coordinates ($r$, $\phi$, $z$) as \cite{forbes2021relevance, green2023optical}:

\begin{align}
    \mathbf{E}_\text{LG}(\mathbf{r}) & =\sum_{k,\ell,p} {\Omega} \biggl[\Bigl(\alpha \mathbf{\hat{x}}+\beta \mathbf{\hat{y}}+\mathbf{\hat{z}}\frac{i}{k}\Bigl\{\alpha\Bigl(\gamma \cos\phi - \frac{i \ell }{r} \sin\phi\Bigr)+\beta \Bigl(\gamma \sin\phi + \frac{i \ell }{r} \cos\phi\Bigr) \Bigr\} \Bigr) \nonumber \\ & \times f_{\mathrm{LG}}\; {a}_{|\ell|,p}\; \mathrm{e}^{i(kz+\ell\phi)} - \mathrm{H.c}\biggr],
\label{eq:2}
\end{align}

and

\begin{align}
    \mathbf{B}_{\mathrm{\text{LG}}}(\mathbf{r}) &=\sum_{k,\ell,p} \frac{{\Omega}}{c} \biggl[\Bigl(\alpha \mathbf{\hat{y}}-\beta \mathbf{\hat{x}}+\mathbf{\hat{z}}\frac{i}{k}\Bigl\{\alpha\Bigl(\gamma \sin\phi + \frac{i \ell }{r} \cos\phi\Bigr)-\beta \Bigl(\gamma \cos\phi - \frac{i \ell }{r} \sin\phi\Bigr) \Bigr\}\Bigr) \nonumber \\
    & \times f_{\mathrm{LG}}\; {a}_{|\ell|,p}\; \mathrm{e}^{i(kz+\ell\phi)} - \mathrm{H.c}\biggr].
\label{eq:3}
\end{align}

In the above equations ${{a}}_{|\ell|,p}$ is the annihilation operator; $\ell$ is the aforementioned topological charge; $p$ is the radial index of Laguerre-Gaussian beams $ p \in \mathbb{N} $. The topological charge has received significantly more attention than the radial index and very often $p=0$ beams are studied. The radial index does have a significant influence on the amplitude distribution, most clearly seen by the fact there exists $p+1$ concentric rings of intensity in a Laguerre-Gaussian mode \cite{andrews2012angular}; $k$ is the wavenumber; $c$ is the speed of light; $\alpha$ and $\beta$ are the, in general, complex Jones coefficients restricted to $|\alpha|^2 + |\beta|^2 =1$;  ${\Omega} = i \left(\hbar ck  / 2\epsilon_0V A_{|\ell|,p}^{2}\right)^{1/2}$ is a normalization constant for the mode expansion where $\epsilon_0$ is the vacuum permittivity; $V$ is the quantization volume; $A_{|\ell|,p}$ is a constant for LG modes \cite{romero2002quantum}; the radial distribution function $f_{\mathrm{LG}}$ of the LG modes at $z=0$ is given by $f_{\mathrm{LG}}(r) = \frac{C^{|{\ell}|}_p}{w_0} \left(\frac{\sqrt{2}r}{w_0}\right)^{|{\ell}|}\exp{\left(\frac{-r^2}{w_0^2}\right)}L_p^{|{\ell}|}$;  $\gamma = \frac{\partial f_{\mathrm{LG}}(r)}{\partial r} = \frac{|{\ell}|}{r}-\frac{2r}{w_0^2}-\frac{4r}{w_0^2}\frac{L^{|{\ell}|+1}_{p-1}}{L^{|{\ell}|}_p}$; $w_0$ is the beam waist, $C^{|\ell|}_p$ is a constant \cite{andrews2012angular} and $L^{|\ell|}_p$ is the associated Laguerre polynomial of order $p$ which has an argument $\bigl[\frac{2r^2}{w_0^2}\bigr]$; $\mathrm{H.c}$ denotes the Hermitian conjugate (the terms attached to the raising operator $a^\dagger$); 

A paraxial beam of light propagating along $z$ has transverse electromagnetic field components in the $xy$-plane, i.e. its polarization state is two-dimensional 2D. For example, the $\hat{\mathbf{x}}$ and $\hat{\mathbf{y}}$-dependent terms in Eq.~\eqref{eq:2} represent the 2D polarization state of the electric field. Such paraxial light, with transverse electromagnetic fields, is well understood and described by the four Stokes parameters. Under focusing, electromagnetic beams become non-paraxial and acquire a longitudinal component in the direction of propagation (the $\hat{\mathbf{z}}$-dependent terms in Eqs.~\eqref{eq:2} and \eqref{eq:3}), the magnitude of which grows with a tighter focus relative to the transverse components \cite{novotny2012principles, adams2018optics}. The fields are now polarized in three dimensions and we refer to the light as being 3D polarized \cite{alonso2023geometric}, described by nine Stokes parameters. 

Let an incident monochromatic Laguerre-Gaussian beam  of $n$ photons propagating along $z$ be focused on to a system of $N$ molecules in the ground state. The initial non-degenerate state is given by $\left| n(k,\eta,\ell,p) \right\rangle \prod_{\xi}^{N} \left| E_o(\xi) \right\rangle$. The final state after the absorption of two photons is an $N$-fold degenerate state corresponding to the $N$ ways of choosing the molecule to be excited, e.g. $\left| (n-2)(k,\eta,\ell,p) \right\rangle \left| E_m(\xi) \right\rangle \prod_{\xi^{'}\neq\xi}^{N} \left| E_o(\xi^{'}) \right\rangle $ \cite{craig1998molecular}. Following the established method of calculating the matrix element $M_{FI}$ for two-photon absorption using second-order time-dependent perturbation theory \cite{power1975matrix, craig1998molecular}:

\begin{align}
     M_{FI} &= \sum_{R} \frac{\bra{F}H_{\mathrm{int}}\ket{R}\bra{R}H_{\mathrm{int}}\ket{I}}{E_{I}-E_{R}} \nonumber \\
     &=\Biggl[f_{\mathrm{LG}}^2\; \mathrm{e}^{2i(kz+\ell\phi)} \sqrt{n(n-1)} \Biggl(\frac{\hbar k} {2{A^2_{|\ell|,p}}V\epsilon_0} \Biggr) \nonumber \\
     &\times
    \Biggl\{\frac{1}{2}ce_ie_j\alpha^{mo}_{ij}(\omega,\omega)+e_ib_jG^{mo}_{ij}(\omega,\omega)+ \frac{1}{2}c^{-1}b_ib_j\chi^{mo}_{ij}(\omega,\omega)\Biggr\}\Biggr],
\label{eq:4}
\end{align}

where the Einstein summation convention for repeated indices is assumed. $H_\text{int}$ is given by Eq.~\eqref{eq:1} and the electromagnetic field operators Eqs.~\eqref{eq:2} and \eqref{eq:3} for a Laguerre-Gaussian beam have been used (a step-by-step derivation is provided in \cite{supplement}). In \eqref{eq:4} a summation is taken over all strongly allowed intermediate states $\ket{R}$ (with molecular energy $E_r$): the selection rules to which this applies are governed by the polarizability tensor $\alpha^{\mathrm{mo}}_{ij}$ and mixed electric-magnetic $G^{\mathrm{mo}}_{ij}$ and purely magnetic $\chi^{\mathrm{mo}}_{ij}$ analogues. The general form of these tensors can be found in refs \cite{salam2009molecular, supplement}. The vector properties of the electromagnetic field, $\mathbf{e}$ and $\mathbf{b}$, (in suffix notation $e_i$ and $b_i$) in Eq.~\eqref{eq:4} are:

\begin{align}
    \mathbf{e} = \alpha \mathbf{\hat{x}}+\beta \mathbf{\hat{y}}+\mathbf{\hat{z}}\frac{i}{k}\Bigl(\alpha\Bigl\{\gamma \cos(\phi) - \frac{i \ell }{r} \sin(\phi)\Bigr\}+\beta \Bigl\{\gamma \sin(\phi) + \frac{i \ell }{r} \cos(\phi)\Bigr\} \Bigr),
\label{eq:5}
\end{align}

and

\begin{align}
    \mathbf{b} = \alpha \mathbf{\hat{y}}-\beta \mathbf{\hat{x}}+\mathbf{\hat{z}}\frac{i}{k}\Bigl(\alpha\Bigl\{\gamma \sin(\phi) + \frac{i \ell }{r} \cos(\phi)\Bigr\}-\beta \Bigl\{\gamma \cos(\phi) - \frac{i \ell }{r} \sin(\phi)\Bigr\} \Bigr).
\label{eq:6}
\end{align}

The TPA rate for an assembly of $N$ absorbing molecules
follows directly from Fermi's golden rule \cite{craig1998molecular}, i.e. the rate of absorption is $\Gamma = N 2\pi |M_{FI}|^2 \rho/\hbar$. The chiroptical contribution to the two-photon absorption rate in an isotropic chiral medium comes from the interference between the electric field coupling to the polarizability tensor $\alpha$ (even parity) and the electromagnetic field coupling to the electric-magnetic analogue $G$ (odd parity): `$\alpha G$' coupling (odd parity) which is only supported by chiral molecules \cite{andrews2018quantum}. This contribution thus contains the fourth rank molecular tensor $\alpha_{ij}\bar{G}_{kl}$ and is determined to be:

\begin{align}
    \Gamma_{\alpha G} = \frac{N\pi\rho}{\hbar}\frac{\bar{I}^2g^{(2)}}{2c^3\epsilon_0^2} \biggl[\text{Re}(e_ie_j\bar{e}_k\bar{b}_l\alpha_{ij}\bar{G}_{kl})\biggr]{f_{\mathrm{LG}}^4}
\label{eq:7}
\end{align}

where we define the mean beam intensity $\bar{I}=\langle n\rangle\hbar c^2k / A^2_{|\ell|,p}V$ and the degree of second-order coherence $g^{(2)}=\langle n(n-1)\rangle / \langle n\rangle^2$; $\rho$ is the density of final states.
The rate as expressed in Eq.~\eqref{eq:7} applies to molecules with fixed orientation with respect to the incoming beam of light. We perform a fourth-rank rotational average \cite{craig1998molecular} on Eq.~\eqref{eq:7} to account for an isotropic ensemble of absorbing chiral molecules (e.g. a solution of freely tumbling chiral molecules):

\begin{align}
    \langle\Gamma_{\alpha G}\rangle &= \text{Re}\frac{N\pi\rho \;\bar{I}^2g^{(2)}}{60\hbar c^3\epsilon_0^2}{f_{\mathrm{LG}}^4} \biggl[(\mathbf{e}\cdot{\mathbf{e}})(\bar{\mathbf{e}}\cdot\bar{\mathbf{b}})\Bigl\{\frac{10}{3}\alpha^{(0)}_{\lambda\lambda}\bar{G}^{(0)}_{\pi \pi} -2\alpha^{(2)}_{\lambda\pi}\bar{G}^{(2)}_{\lambda\pi}\Bigr\}  +  (\mathbf{e}\cdot\bar{\mathbf{e}})(\mathbf{e}\cdot\bar{\mathbf{b}})6\alpha^{(2)}_{\lambda\pi}\bar{G}^{(2)}_{\lambda\pi} \biggr].
\label{eq:8}
\end{align}

This is the final form of the rate expression which follows from the fourth-rank rotational average followed by a decomposition into irreducible components \cite{craig1998molecular,forbes2022two}: scalar (weight 0), denoted by a superscript (0), and traceless (weight 2) by superscript (2). The usefulness of analysing the result in terms of irreducible components is that selection rules for TPA can be readily derived for a given molecular point group (we make use of this later) as the weight of an irreducible tensor responsible for a given transition correlates to the angular momentum of the photon(s) involved in the transition. Expanding out the scalar products in Eq.~\eqref{eq:8} leads to a large expression due to its complete generality with respect to the input beam 2D state of polarization, i.e. the values of $\alpha$ and $\beta$. As stated, we are concentrating on 2D linearly polarized states in this paper - the Jones vector coefficients $\alpha$ and $\beta$ are real - which gives the following for two important cases:

2D $x$-polarized input beam ($\alpha=1$,$\beta=0$):

\begin{align}
        (\mathbf{e}\cdot{\mathbf{e}})(\bar{\mathbf{e}}\cdot\bar{\mathbf{b}}) & =\frac{i\ell\gamma}{k^2r}\cos2\phi+\frac{i}{k^4}\Bigl[\Bigl(\frac{\ell\gamma^3}{r}+\frac{\ell^3\gamma}{r^3}\Bigr)(\cos^4\phi-\cos^2\phi)\Bigr]
    \label{eq:9}
    \end{align}

\begin{align}
        (\mathbf{e}\cdot\bar{\mathbf{e}})(\mathbf{e}\cdot\bar{\mathbf{b}})=-\frac{i\ell\gamma}{k^2r}+\frac{i}{k^4}\Bigl[\Bigl(\frac{\ell\gamma^3}{r}+\frac{\ell^3\gamma}{r^3}\Bigr)(\cos^4\phi-\cos^2\phi)\Bigr]
    \label{eq:10}
\end{align}

2D $y$-polarized input beam ($\alpha=0$,$\beta=1$):

\begin{align}
        (\mathbf{e}\cdot{\mathbf{e}})(\bar{\mathbf{e}}\cdot\bar{\mathbf{b}})&=-\frac{i\ell\gamma}{k^2r}\cos2\phi+\frac{i}{k^4}\Bigl[\Bigl(\frac{\ell\gamma^3}{r}+\frac{\ell^3\gamma}{r^3}\Bigr)(\cos^4\phi-\cos^2\phi)\Bigr]
    \label{eq:11}
\end{align}

\begin{align}
        (\mathbf{e}\cdot\bar{\mathbf{e}})(\mathbf{e}\cdot\bar{\mathbf{b}})=-\frac{i\ell\gamma}{k^2r}+\frac{i}{k^4}\Bigl[\Bigl(\frac{\ell\gamma^3}{r}+\frac{\ell^3\gamma}{r^3}\Bigr)(\cos^4\phi-\cos^2\phi)\Bigr]
    \label{eq:12}
\end{align}

The terms dependent on $1/k^2d^2$, where $d$ is either $r$ or $w_0$, come from the interference of the transverse and longitudinal components of the electromagnetic field; those dependent on $1/k^4d^4$ stem purely from the longitudinal field components (and are thus weaker in general compared to the former). 

First and foremost it is obvious that for 2D linearly polarized non-vortex beams $\ell = 0$, i.e. a Gaussian beam or a plane-wave, then Eq.~\eqref{eq:8} is zero: for such `unstructured' excitation sources a degree of ellipticity in the 2D polarization is required to exhibit optical activity. Moreover, in theoretical accounts the traditional electromagnetic field source would be a circularly polarized plane wave (i.e. a paraxial electromagnetic field with no longitudinal components): $\mathbf{e}=\hat{\mathbf{x}}\pm i\hat{\mathbf{y}}$ and $\mathbf{b}=\hat{\mathbf{y}}\mp i\hat{\mathbf{x}}$. Clearly in such a case $\mathbf{e}\cdot\mathbf{e}=\bar{\mathbf{e}}\cdot\bar{\mathbf{b}}=0$ and these methods cannot access the structural information provided by the $\alpha^{(0)}_{\lambda\lambda}\bar{G}^{(0)}_{\pi \pi}$ tensor: in contrast a focused optical vortex provides unique access to the information in both $\alpha^{(2)}_{\lambda\pi}\bar{G}^{(2)}_{\lambda\pi}$ and $\alpha^{(0)}_{\lambda\lambda}\bar{G}^{(0)}_{\pi \pi}$. In fact it is quite clear that for a single beam, only a focused one with longitudinal components has the potential to fulfill both $(\mathbf{e}\cdot{\mathbf{e}}) \neq 0$ 
 and $(\bar{\mathbf{e}}\cdot\bar{\mathbf{b}})\neq0$ necessary to engage $\alpha^{(0)}_{\lambda\lambda}\bar{G}^{(0)}_{\pi \pi}$. A 2D circularly polarized Gaussian beam which is focused can achieve this but the coupling to $\alpha^{(0)}_{\lambda\lambda}\bar{G}^{(0)}_{\pi \pi}$ is of order $1/k^4w_0^4$, unlike vortex beams which is $1/k^2d^2$ (see \cite{supplement} for details). In single beam TPA with a plane-wave source both weight 0 and weight 2 are in general allowed \cite{craig1998molecular}, the former when the input beam is linearly polarized (expectation value of angular momentum is zero) and the latter when the input beam is circularly polarized (two photons with an angular momentum of $\sigma \hbar$ each). In TPA-CD weight 0 is completely inaccessible due to the necessity of circularly polarized photons: this is not the case in our TPA-VD which manifests with linearly-polarized light. Moreover, as shown in \cite{forbes2022two, supplement}, a circularly-polarized Gaussian beam can access the weight 0 tensor but only if focused (i.e. non-zero longitudinal field components). This result completely tallies with the well-known spin-orbit angular momentum conversion which takes place under the focusing of a circularly-polarized Gaussian beam \cite{bliokh2015spin}. The beam before focusing has an expectation value of spin angular momentum $\sigma \hbar$ per photon; in the focal plane this expectation value of spin decreases (the tighter the focus, the more it does so) and the orbital angular momentum increases in order to conserve total angular momentum. Under dipole coupling it is the spin-angular momentum of light which couples to the electronic degrees of freedom in molecules, thus under focusing the weight 0 tensor can drive transitions. These results are summarized in Table \ref{tab1}.
 
\begin{table}[]
    \centering
    \caption {Allowed weights in TPA for vortex $\ell\neq0$ and fundamental Gaussian beams $\ell=0$ with either linear or circular (elliptical) polarization under different focusing conditions.} \label{tab1} 
\begin{tabular}{l*{6}{c}r}
      & Paraxial & Nonparaxial  \\
\hline
$\sigma =0, \ell=0$ & - & -  \\
$\sigma\neq0, \ell=0$ & $\alpha^{(2)}_{\lambda\pi}\bar{G}^{(2)}_{\lambda\pi}$ & $\alpha^{(0)}_{\lambda\lambda}\bar{G}^{(0)}_{\pi \pi}$, $\alpha^{(2)}_{\lambda\pi}\bar{G}^{(2)}_{\lambda\pi}$   \\
$\sigma=0, \ell\neq0$           & - & $\alpha^{(0)}_{\lambda\lambda}\bar{G}^{(0)}_{\pi \pi}$, $\alpha^{(2)}_{\lambda\pi}\bar{G}^{(2)}_{\lambda\pi}$   \\
$\sigma\neq0, \ell\neq0$    & $\alpha^{(2)}_{\lambda\pi}\bar{G}^{(2)}_{\lambda\pi}$ & $\alpha^{(0)}_{\lambda\lambda}\bar{G}^{(0)}_{\pi \pi}$, $\alpha^{(2)}_{\lambda\pi}\bar{G}^{(2)}_{\lambda\pi}$   \\
\end{tabular} \\
\end{table}

\begin{figure}[H]
    \centering
    \begin{minipage}{0.49\textwidth}
    \centering
    \includegraphics[width=1\linewidth]{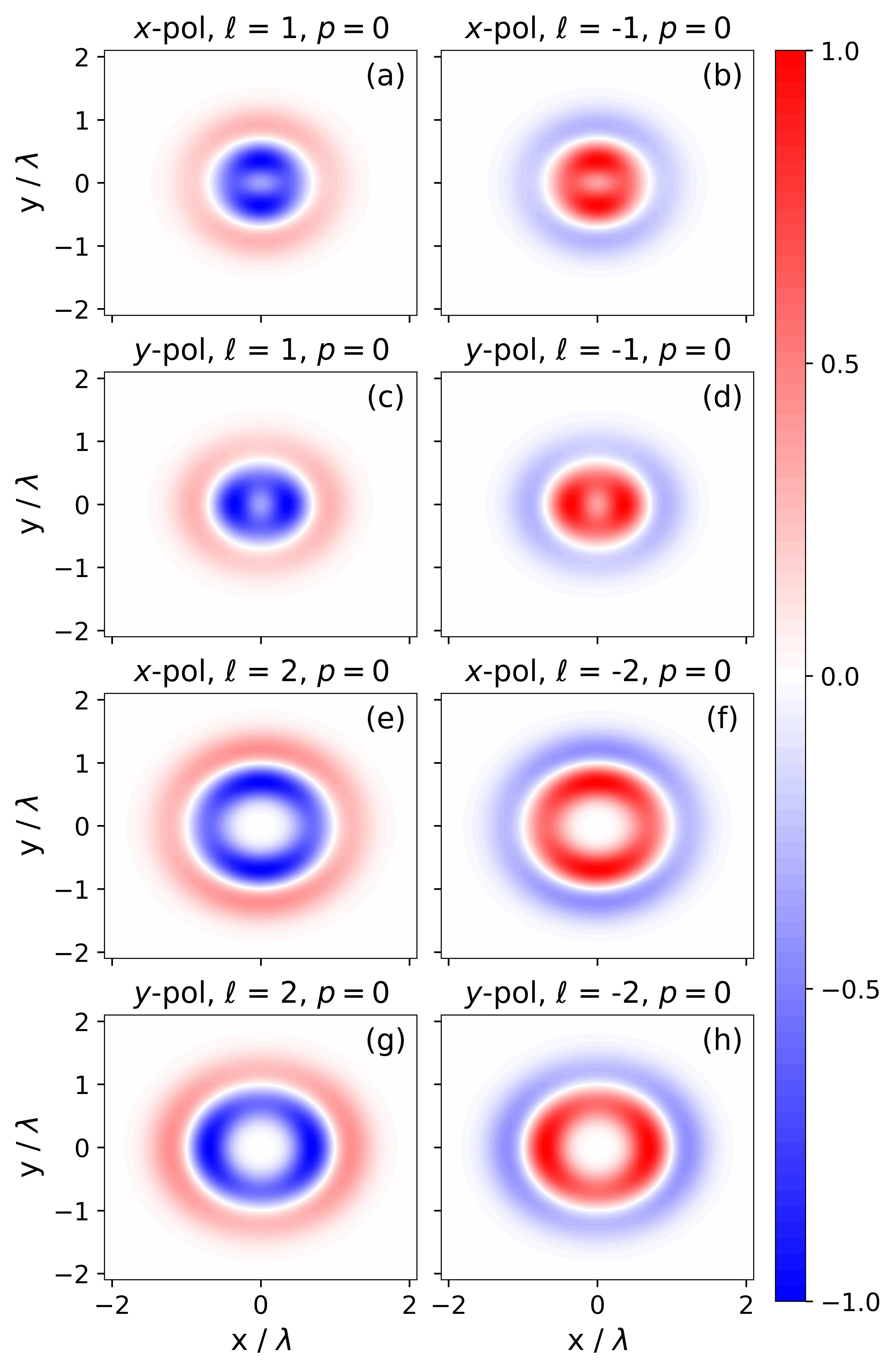}
    \caption{In (a)-(h): the normalised absorption rate Eq. \eqref{eq:8} is plotted in the xy-plane at $z=0$ for left and right vortex handedness; $|\ell|=1$ (a)-(d), $2$ (e)-(h); $p=0$, $w_0= \lambda = 7.29\times10^{-9}$ m; photons x-polarized in (a),(b),(e),(f) and y-polarized in (c),(d),(g),(h). In all Figures red indicates an increased rate of absorption, blue corresponds to a decreased absorption rate.}
    \label{fig2}   
    \end{minipage}
    \hfill
    \begin{minipage}{0.49\textwidth}
    \centering
    \includegraphics[width=1\linewidth]{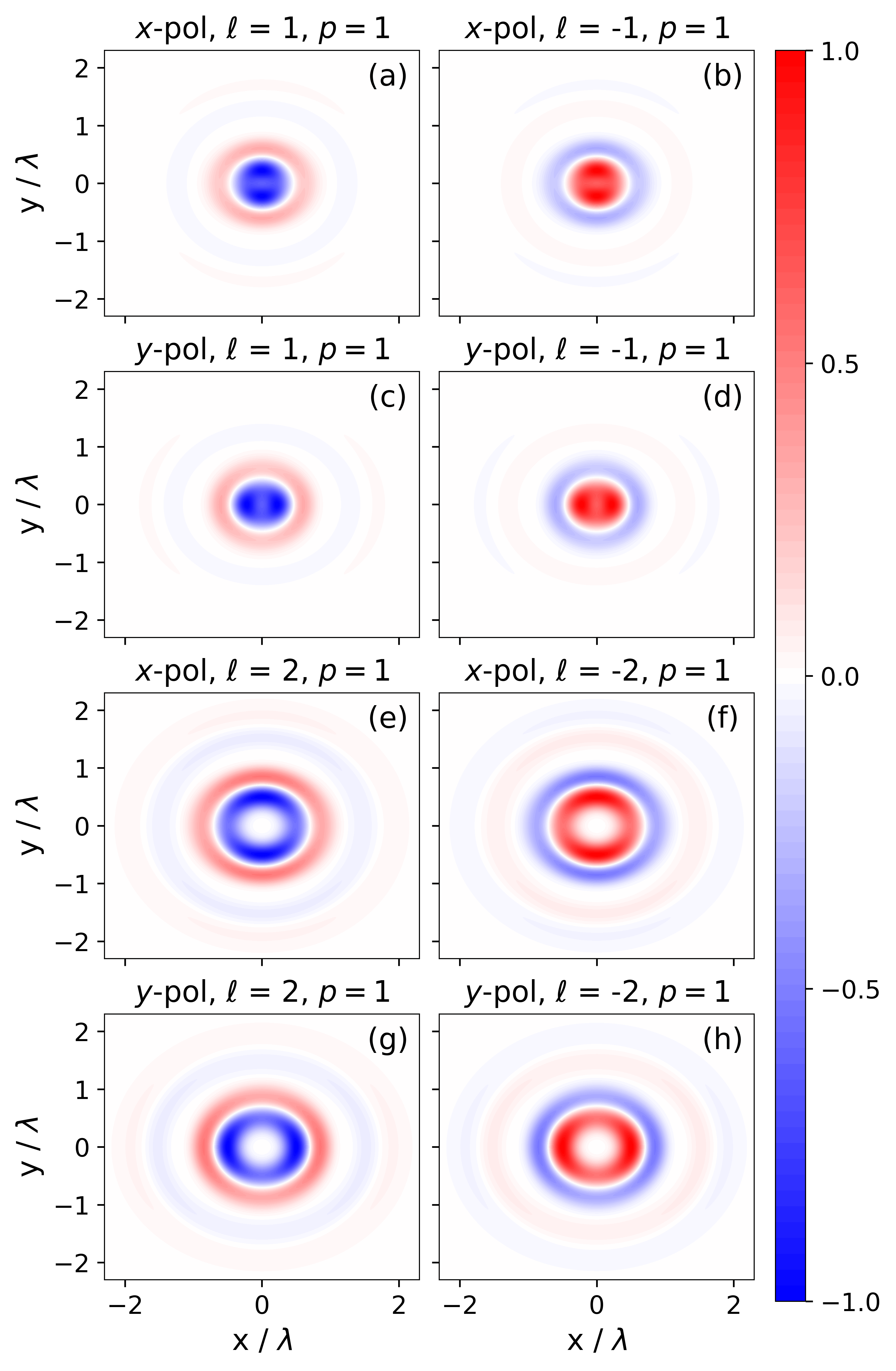}
    \caption{In (a)-(h): the normalised absorption rate Eq. \eqref{eq:8} is plotted in the xy-plane at $z=0$ for left and right vortex handedness; $|\ell|=1$ (a)-(d), $2$ (e)-(h); $p=1$, $w_0= \lambda = 7.29\times10^{-9}$ m; photons x-polarized in (a),(b),(e),(f) and y-polarized in (c),(d),(g),(h).}  
    \label{fig3}  
    \end{minipage}
\end{figure}

To visualize the spatial distribution of nonlinear VD for input 2D linearly polarized photons the normalised two-photon absorption rate Eq.~\eqref{eq:8} is plotted in the $xy$-plane at $z=0$ (the point of maximum spatial confinement at which $w(z) = w_0$) in Figure \ref{fig2} and Figure \ref{fig3} under the assumption $\alpha^{(0)}_{\lambda\lambda}\bar{G}^{(0)}_{\pi \pi} =\alpha^{(2)}_{\lambda\pi}\bar{G}^{(2)}_{\lambda\pi}$ (see below for further discussion of this). The results for 2D circularly polarized beams can be found in \cite{supplement}.
 
Immediately evident is the result that the absorption of vortex photons by chiral molecules is dependent upon the sign of $\ell$, i.e. the wavefront chirality: there is no need for any degree of ellipticity in the input beam (i.e. circular/elliptical polarization) to generate optical activity, specifically a new chiroptical form of dichroism. In other words nonlinear VD can be defined by the equation $\langle\Gamma_{|\ell|}^{(\sigma=0)}\rangle - \langle\Gamma_{-|\ell|}^{(\sigma=0)}\rangle \neq 0$. An extraordinary property of nonlinear VD is that the spatial distribution of the absorption rate is dependent on the orientation of the 2D state of linear polarization of the input beam (i.e. the longitude position - or azimuth - on the Poincar\'e sphere) when vortex handedness $\ell$ remains constant. This is a new and unexpected effect of vortex chirality that has only become apparent in the analysis of VD in a nonlinear absorption process. We are not aware of any other optical activity which depends upon the orientation of the state of linear polarization of an input beam. Although surprising in this context, nonlinear light-matter interactions have long been known to exhibit different behaviour with respect to the optical state of polarization compared to the linear regime \cite{andrews1977polarization, andrews1981polarization}. It is clear that the first term on the right-hand side of Eqs.~\eqref{eq:9} and \eqref{eq:11} are predominantly responsible for this rotation of the nonlinear VD rate of absorption, i.e. for a given value of $\ell$, 2D-$x$-polarized LG photons have a $+\cos2\phi$ dependence whereas 2D-$y$-polarized LG photons have a $-\cos2\phi$ dependence.  

In the cases of $\ell=\pm1$ there is small on-axis absorption of the light at the centre of the focused beam, even though before focusing the input beam had an intensity null at the centre. The fact that optical vortices under focusing have a non-zero central intensity is well-known \cite{bliokh2015spin}. From Eqs.~\eqref{eq:9}-\eqref{eq:12} the pure longitudinal field terms dependent on $1/k^4$ have terms with a $1/r^4$ dependence, which for $|\ell|=1$ are multiplied by a factor of $r^4$ stemming from $f_\text{LG}^4$ in the full rate Eq.~\eqref{eq:8} giving contributions to the rate whose only $r$-dependence comes from the Gaussian function, which is thus responsible for the on-axis nonlinear VD. 

It is important to notice that each absorption rate profile takes on both positive and negative signs. This is unlike traditional CD signals where circular polarization produces an intensity signal of uniform sign for a given handedness (see also supplementary information \cite{supplement} for the nonlinear circular-vortex dichroism case). This is a characteristic of VD also showcased in linear interactions \cite{ovdicp} which indicates that the size (and number) of the chiral particles in the assembly is important in all instances of VD which is reflective of the scale-dependent nature of VD. Figure \ref{fig3} highlights the role that increasing the radial index $p$ has on the spatial distribution. Analogous to linear VD it produces $2p+2$ rings of absorbing regions, however unlike linear VD these rings also rotate with the azimuth (in fact the outer `ring' of the $\ell=1$ cases highlight this rotation clearly). Another result of increasing $p$ is that the inner rings of absorption become smaller (i.e. increasing $p$ generates a tighter focus for the central bright rings), and the on-axis absorption increases \cite{forbes2022two}. This mirrors previous results in the linear regime \cite{green2023optical}. Increasing the radial index $p$ increases the gradient of the transverse profile of the field (by adding in $p+1$ rings of intensity). Longitudinal field components are proportional to the gradient of the transverse field, which increasing $p$ increases, leading to larger on-axis absorption. 

\begin{figure*}
    \includegraphics[]{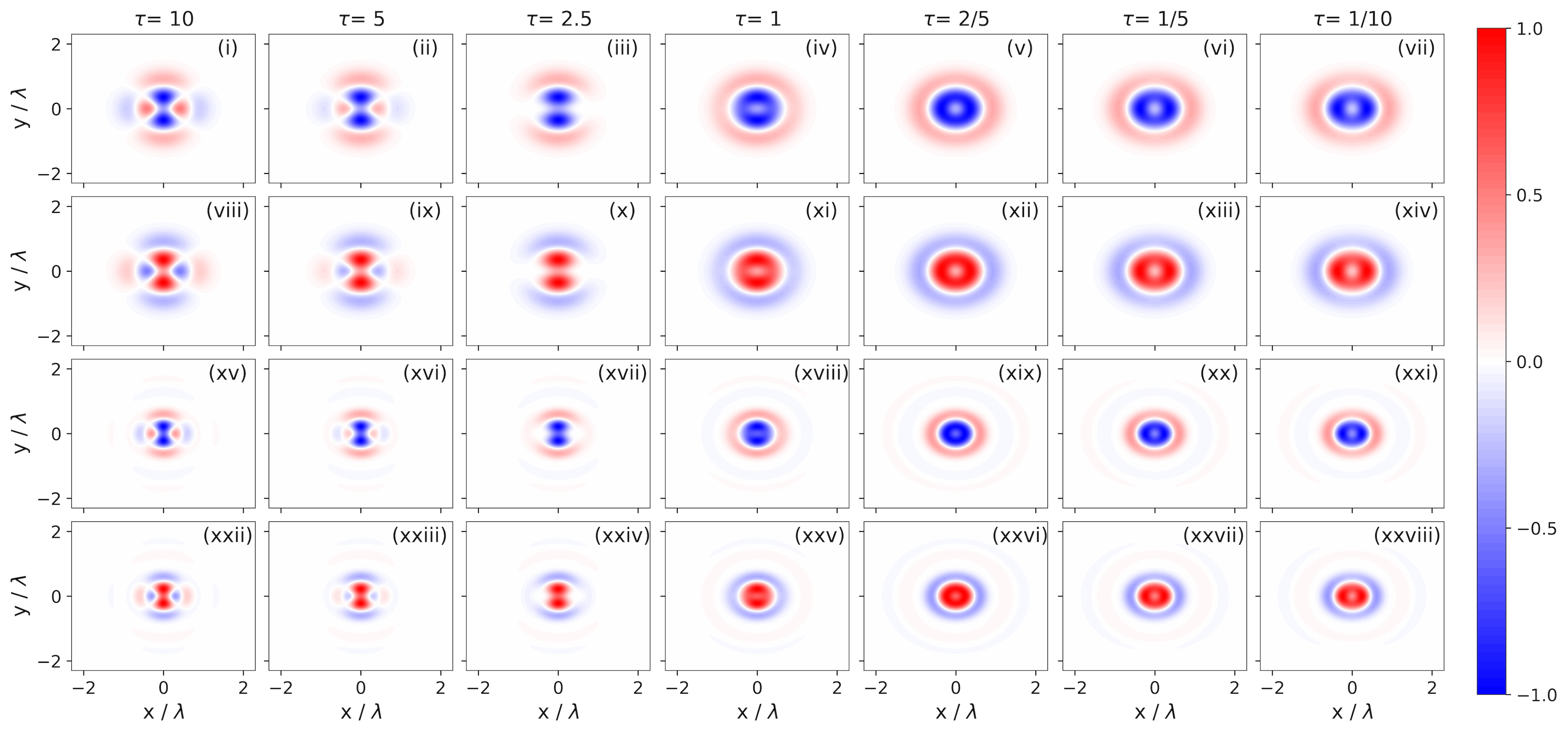}
    \caption{the normalised absorption rate Eq. \eqref{eq:8} is plotted in the xy-plane at $z=0$ as a function of $\tau = \alpha^{(0)}_{\lambda\lambda}\bar{G}^{(0)}_{\pi \pi}/ \alpha^{(2)}_{\lambda\pi}\bar{G}^{(2)}_{\lambda\pi}$ for $\ell=1, p=0$ (i)-(vii); $\ell=-1, p=0$ (viii)-(xiv); $\ell=1, p=1$ (xv)-(xxi); and $\ell=-1, p=1$ (xxii)-(xxviii). In all plots the input beam is 2D $x$-polarized ($\alpha = 1, \beta = 0)$ and $w_0= \lambda = 7.29\times10^{-9}$ m}
    \label{fig4}
\end{figure*}

Figures \ref{fig2} and \ref{fig3} are for the specific case of $\alpha^{(0)}_{\lambda\lambda}\bar{G}^{(0)}_{\pi \pi} =\alpha^{(2)}_{\lambda\pi}\bar{G}^{(2)}_{\lambda\pi}$ to give an indicative representation of the spatial distribution of NVD. The rate Eq.~\eqref{eq:8} is of course strongly influenced by the values of the polarizability tensors $\alpha$ and $G$ and the most general situation is $\alpha^{(0)}_{\lambda\lambda}\bar{G}^{(0)}_{\pi \pi} 
 \neq \alpha^{(2)}_{\lambda\pi}\bar{G}^{(2)}_{\lambda\pi} \neq 0$. Quantitative values for the molecular polarizability tensors for specific chiral molecules require quantum chemistry calculations far beyond the scope of this study. Nonetheless, Figure \ref{fig4} gives an example of how the ratio $\tau = \alpha^{(0)}_{\lambda\lambda}\bar{G}^{(0)}_{\pi \pi}/ \alpha^{(2)}_{\lambda\pi}\bar{G}^{(2)}_{\lambda\pi}$ significantly influences the spatial distribution of NVD for a chiral molecule making transitions to a state where both weight 0 and weight 2 are allowed, e.g. the transition to an $A$ state in a chiral molecule belonging to $D_2$. What we see is that the ratio of $\tau = \alpha^{(0)}_{\lambda\lambda}\bar{G}^{(0)}_{\pi \pi}/ \alpha^{(2)}_{\lambda\pi}\bar{G}^{(2)}_{\lambda\pi}$, which is a rich source of molecular information and allows the symmetry of electronic excited states to be established, significantly influences the spatial distribution of NVD. Alternatively put, the spatial distribution of absorption gives a detailed insight into molecular structure and excited state symmetry through its strong dependence on $\tau$.

\begin{figure}[H]
    \centering
    \begin{minipage}{0.49\textwidth}
    \centering
    \includegraphics[width=1\linewidth]{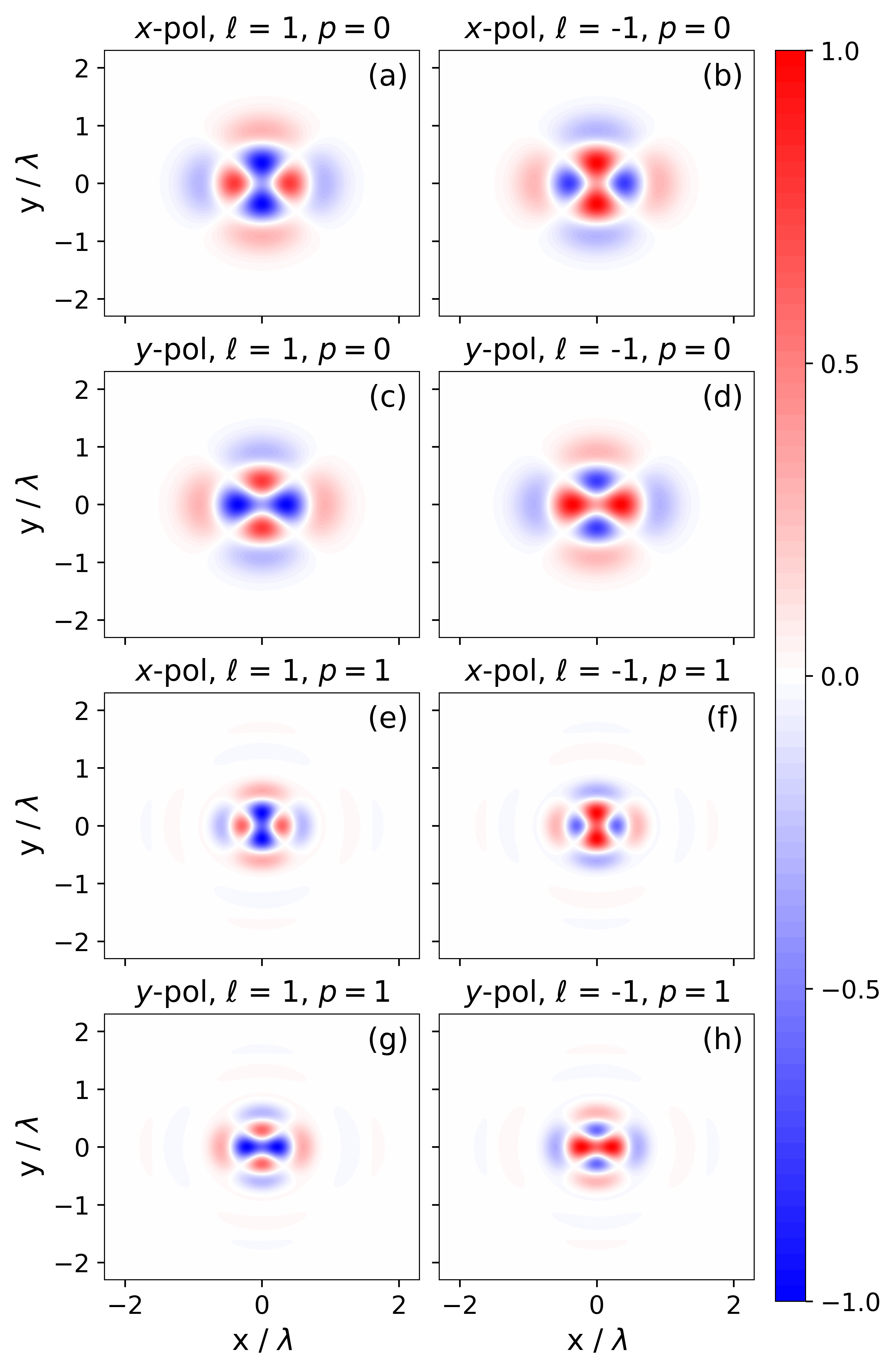}
    \caption{In (a)-(h): the normalised absorption rate Eq. \eqref{eq:8} is plotted in the xy-plane at $z=0$ for left and right vortex handedness; $|\ell|=1, p =0$ (a)-(d), $|\ell|=1,p=1$ (e)-(h); $w_0= \lambda = 7.29\times10^{-9}$ m; photons x-polarized in (a),(b),(e),(f) and y-polarized in (c),(d),(g),(h). In all plots $\alpha^{(0)}_{\lambda\pi}\bar{G}^{(0)}_{\lambda\pi} = 1$, $\alpha^{(2)}_{\lambda\pi}\bar{G}^{(2)}_{\lambda\pi} = 0$}
    \label{fig5}   
    \end{minipage}
    \hfill
    \begin{minipage}{0.49\textwidth}
    \centering
    \includegraphics[width=1\linewidth]{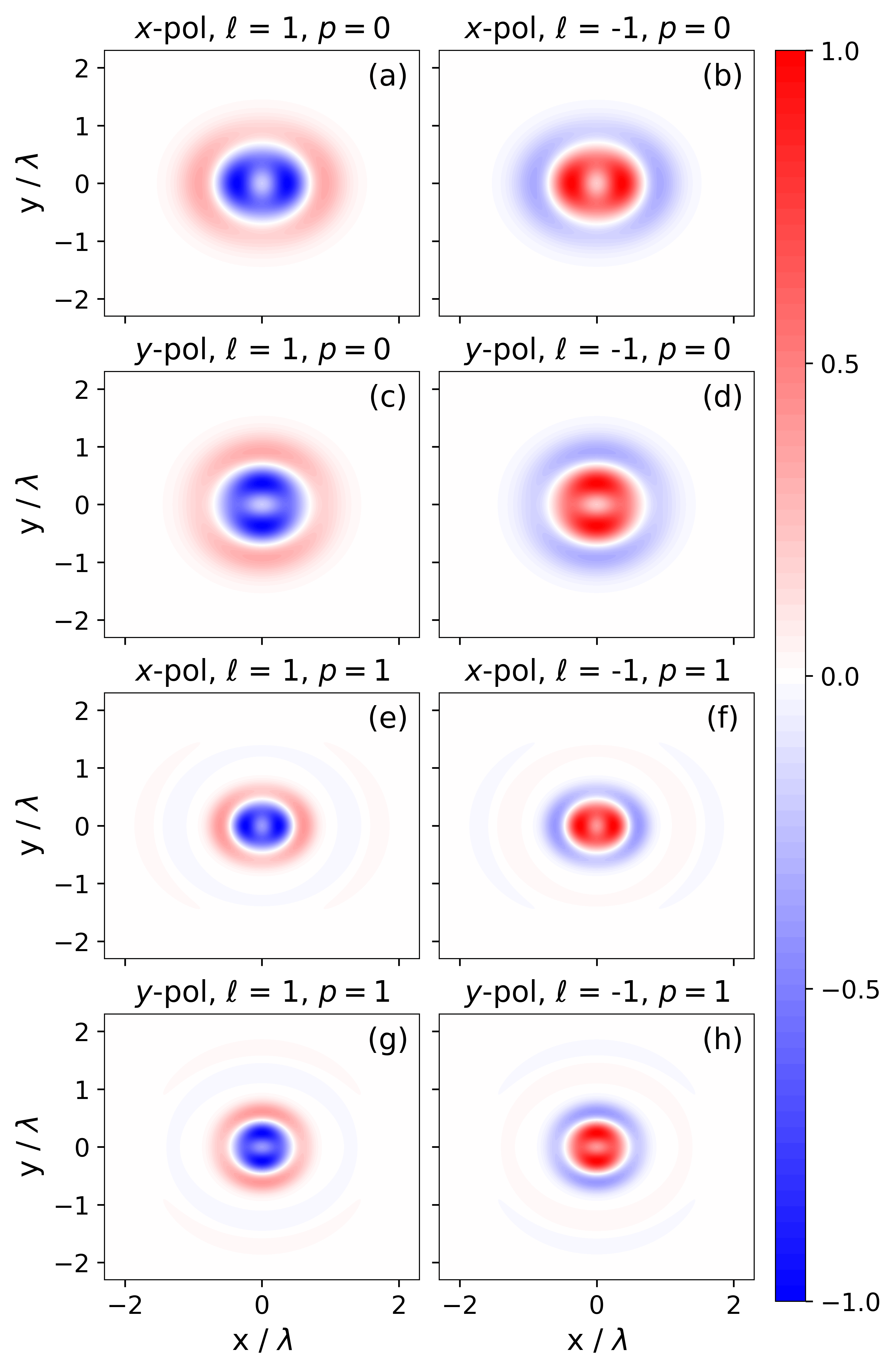}
    \caption{In (a)-(h): the normalised absorption rate Eq. \eqref{eq:8} is plotted in the xy-plane at $z=0$ for left and right vortex handedness; $|\ell|=1, p =0$ (a)-(d), $|\ell|=1,p=1$ (e)-(h); $w_0= \lambda = 7.29\times10^{-9}$ m; photons x-polarized in (a),(b),(e),(f) and y-polarized in (c),(d),(g),(h). In all plots $\alpha^{(0)}_{\lambda\pi}\bar{G}^{(0)}_{\lambda\pi} = 0$, $\alpha^{(2)}_{\lambda\pi}\bar{G}^{(2)}_{\lambda\pi} = 1$} 
    \label{fig6}  
    \end{minipage}
\end{figure}

Moreover, because certain point groups have irreducible representations which allow only weight 0 or only weight 2 contributions \cite{andrews1989three, andrews1990symmetry, andrews2002optical} there will be molecular systems which exhibit $\alpha^{(0)}_{\lambda\lambda}\bar{G}^{(0)}_{\pi \pi} = 0, \alpha^{(2)}_{\lambda\pi}\bar{G}^{(2)}_{\lambda\pi} \neq 0$ and $\alpha^{(0)}_{\lambda\lambda}\bar{G}^{(0)}_{\pi \pi} \neq 0, \alpha^{(2)}_{\lambda\pi}\bar{G}^{(2)}_{\lambda\pi} = 0$. This behaviour, which essentially represents the ratio $\tau$ at the extremes of 0 and $\infty$, is presented in Figure \ref{fig5} and Figure \ref{fig6}. As mentioned, a chiral molecule belonging to $D_2$ making a transition to an $A$ state allows both weight 0 and weight 2 for rank 2 tensors $\alpha^{(0)}_{\lambda\lambda}\bar{G}^{(0)}_{\pi \pi} \neq 0, \alpha^{(2)}_{\lambda\pi}\bar{G}^{(2)}_{\lambda\pi} \neq 0$; however a transition to $B_{1,2,3}$ is only allowed by weight 1 and weight 2 rank 2 tensors $\alpha^{(0)}_{\lambda\lambda}\bar{G}^{(0)}_{\pi \pi} = 0, \alpha^{(2)}_{\lambda\pi}\bar{G}^{(2)}_{\lambda\pi} \neq 0$. Of course, with recourse to character tables such combinations can be found across different point groups and irreducible representations, the key physics however being that the behaviour exhibited in Figs. \ref{fig4}-\ref{fig6} should be readily observed and moreover offer insights into chiral molecular structure. Interestingly, chiral molecules belonging to point groups $T$, $O$, and $I$ allow for solely weight 0 and solely weight 2 contributions. 

\section{Discussion}

The standard method of nonlinear chiroptical spectroscopy based on absorption is two-photon circular dichroism: the differential absorption of circularly polarized photons by chiral molecules. The first theoretical accounts of TPA-CD came in 1975 from Power \cite{power1975two} and Tinoco \cite{tinoco1975two}, followed two decades later by a fluorescence-detected experimental realization in 1995 \cite{gunde1995fluorescence}. A number of studies have followed \cite{jansik2005response, de2008synchronized, jansik2008strong, toro2010two, savoini2012circular, diaz2015two, vesga2018two}, including the first true experimental observation due to the invention of the double L-scan technique \cite{de2008synchronized}. The interest in the method is predominantly twofold: single-photon CD is typically observed in the near and far UV region whereas TPA-CD uses shorter wavelengths which has distinct advantages when studying biological systems and drugs that are soluble in organic solvents. Our TPA-VD mechanism represents the latest addition to a series of new and powerful methods of elucidating chiral molecular structure underpinned by absorption. Clearly TPA-VD is significantly more tailorable than traditional TPA-CD due to the ability to select the values of $\ell$ and $p$, as well as being highly dependent on the input beam polarization state, even manifesting for linearly-polarized beams $\sigma = 0$. Unique attributes of two-photon VD compared to traditional two-photon CD include the fact the technique works with linearly polarized beams, the differential effect stemming form the sign of $\ell$ rather than $\sigma$; it is proportional to $\ell$ which can be increased in magnitude unlike $\sigma$; there is the additional beam parameter $p$ which influences the absorption rate; the polarizability tensors which inform us of the molecular structure coupling to the structured beam in such a way that produces highly distinct spatial distributions of absorption; and finally the use of a focused vortex enables access to structural information - $\alpha^{(0)}_{\lambda\lambda}\bar{G}^{(0)}_{\pi \pi}$ - not accessible to traditional plane-wave (or paraxial) excitation. 

As a form of nonlinear optical activity, TPA dichroism based techniques are only a very small subset of possibilities. Nonlinear optical activity \cite{fischer2005nonlinear} more generally includes techniques based on harmonic and sum-frequency generation, hyper-Rayleigh and hyper-Raman scattering as well as an array of techniques in the ultrafast regime \cite{habibovic2024application}. Some cutting-edge nonlinear optical activity techniques reported recently include hyper-Rayleigh optical activity in both chiral molecules \cite{verreault2019hyper} and nanostructures \cite{collins2019first}, hyper-Raman optical activity in plasmonic-molecular composites \cite{jones2024chirality}, and third harmonic Mie and Rayleigh optical activity \cite{ohnoutek2021optical, ohnoutek2022third}. The first study of vortex chirality engaging through nonlinear light-matter interactions was the hyper-Rayleigh and hyper-Raman optical activity of optical vortex beams \cite{forbes2020nonlinear}. That study was limited to the electric-dipole electric-quadrupole contributions where it was concluded that photons possessing a higher OAM (larger value of $\ell$) produce a larger hyper-Rayleigh and hyper-Raman scattering differential.  Pioneering experiments by Rouxel et al. \cite{rouxel2022hard} observed significantly increased chiroptical signals in nonlinear optical activity of an isotropic ensemble of chiral organometallic compounds subject to a linearly polarized vortex beam in the hard X-ray regime. Similarly, the Bhardwaj group has recently utilized nonlinear interactions of vortex beams with both chiral and achiral molecules in pursuit of new spectroscopic techniques \cite{begin2023nonlinear, jain2023helical, jain2024intrinsic}. 

\section{Conclusion}

Here we have shown that two-photon absorption of focused optical vortex beams leads to a novel form of nonlinear optical activity in chiral molecules. Chiroptical signals can be produced even when the beam is linearly polarized before focusing, the differential absorption being determined by the sign of $\ell$ of the vortex beam. It is striking that the spatial distribution of absorption is significantly influenced by the orientation of the state of linear polarization, a phenomenon we are not aware of occurring in any other form of optical activity previously studied. Moreover, the use of a focused vortex beam allows access to molecular structural information not accessible with current traditional methods based on paraxial beams or plane-wave excitation, and is a unique tool to establish the symmetry of excited electronic states. Extending the application of focused optical vortex beams into other forms of nonlinear optical activity is certain to produce further spectroscopic techniques with distinct advantages and novelty compared to existing methods.

\begin{acknowledgement}

David L. Andrews is thanked for lending expert insight into the symmetry properties of the polarizability tensors. Dale Green is thanked for aiding the construction of Figure 1.

\end{acknowledgement}

\begin{suppinfo}

The supporting information \cite{supplement} contains the a step-by-step derivation of Eq.~\eqref{eq:8}; a full analysis on the case of 2D-circularly polarized beams; the derivation of the E1E2 contribution to NVD.

\end{suppinfo}


\bibliography{Refs.bib}

\end{document}